\documentclass[final,3p,times,twocolumn,authoryear]{elsarticle}

\usepackage{amssymb}
\usepackage{amsmath}
\usepackage{bm}

\usepackage{booktabs}
\usepackage{multirow}
\usepackage{hyperref}
\usepackage{xcolor}

\biboptions{sort}

\journal{Medical Image Analysis}

\begin{document}

\begin{frontmatter}



\title{Bayesian Posterior Sampling for Synthetic Shape Generation of Heart Valves}


\author[me]{Vijay K. Dubey} 
\author[bme]{Sumedh Seetharam}
\author[ase]{Nikos Manthatis}
\author[me]{Collin E. Haese} 
\author[ase]{L. River Spencer}
\author[uksh]{Jakob Christoph Voran}
\author[asklepios]{Marnie Goldmann}
\author[asklepios]{Felix Kreidel}
\author[issam]{Issam Moussa}
\author[penn]{Alison Pouch}
\author[ase,bme,oden]{Manuel K. Rausch} 
\author[ase,oden]{Jan Fuhg} 


\affiliation[penn]{organization={Penn Center for Biomedical Image Computing and Analysis (CBICA), Perelman School of Medicine, University of Pennsylvania},
            city={Philadelphia},
            postcode={19104}, 
            state={Pennsylvania},
            country={U.S.}}

\affiliation[issam]{organization={Carle Illinois College of Medicine, University of Illinois at Urbana-Champaign},
            city={Urbana},
            postcode={61801}, 
            state={Illinois},
            country={U.S.}}

\affiliation[me]{organization={Walker Department of Mechanical Engineering, The University of Texas at Austin},
            city={Austin},
            postcode={78712}, 
            state={Texas},
            country={U.S.}}
            
\affiliation[ase]{organization={Department of Aerospace Engineering \& Engineering Mechanics, The University of Texas at Austin},
	city={Austin},
	postcode={78712}, 
	state={Texas},
	country={U.S.}}

\affiliation[asklepios]{organization={Department of Cardiology, Asklepios Hospital},
	city={Harburg},
	country={Germany}}

\affiliation[uksh]{organization={Department of Internal Medicine III, Cardiology and Critical Care, University Hospital Schleswig-Holstein},
	city={Kiel},
	country={Germany}}
	
\affiliation[bme]{organization={Department of Biomedical Engineering, The University of Texas at Austin},
	city={Austin},
	postcode={78712}, 
	state={Texas},
	country={U.S.}}

\affiliation[oden]{organization={The Oden Institute of Computational Science and Engineering, The University of Texas at Austin},
	city={Austin},
	postcode={78712}, 
	state={Texas},
	country={U.S.}}

\begin{abstract}
Statistical shape models (SSMs) for heart valves commonly rely on principal component analysis (PCA). They are used to support downstream tasks, including \textit{in silico} modeling, morphological analysis, and interventional planning. However, PCA-based SSMs lack a mechanism for conditional shape generation, i.e., they can create non-physical shapes and perform poorly in low-data regimes (<20 shapes). 
To overcome these problems, we propose instead a Bayesian posterior sampling framework to generate valve shapes from a posterior estimate. The prior relies on a Gaussian mixture model with data-driven mixture modes. The likelihood estimate is obtained through a classifier trained to distinguish valid from invalid regions in the compact proper orthogonal decomposition (POD) coefficient space.
We verify the framework on a model problem and validate it on parametrically constructed aortic valve datasets. Thereby, we demonstrate that our method captures multiple modes, respects decision boundaries in shape space, and outperforms PCA-based SSMs in low-data regimes. We also characterize the framework's performance as a function of dataset size, identifying where diminishing returns arise for the proposed generative shape model.
Finally, we apply the framework to a cohort of ten three-dimensional transesophageal echocardiography images of adult human tricuspid valves. We first segment images to extract shapes, then generate a set of physiologically plausible new shapes. We demonstrate downstream applications for both valves, including \textit{in silico} modeling of valve mechanics and synthetic image-mask creation to augment limited datasets.
The proposed approach bootstraps building image-mask datasets more efficiently than PCA-based SSMs. Although demonstrated only for the aortic and tricuspid valves, the methodology is broadly applicable to all valves.
\end{abstract}

\begin{graphicalabstract}
\includegraphics[width=\textwidth]{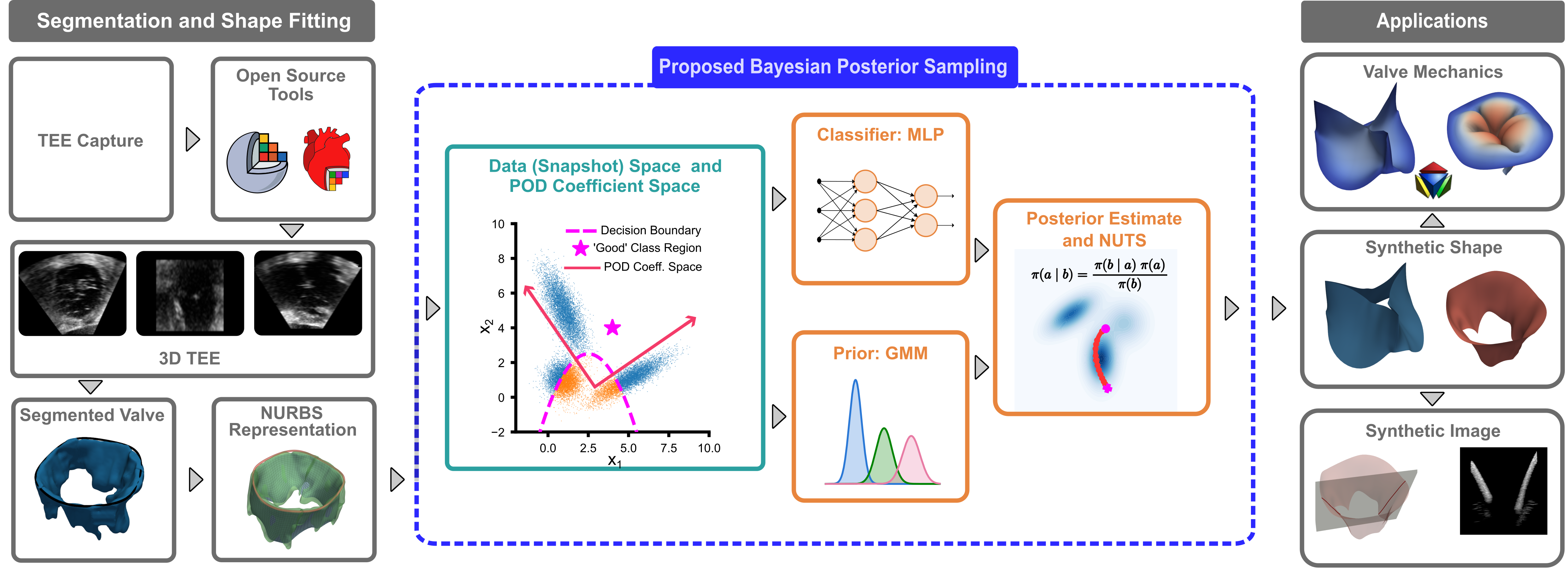}
\end{graphicalabstract}

\begin{highlights}
\item Traditional PCA-based SSMs fail to capture multimodality and non-normality and have no native mechanism for conditional shape generation. Our proposed framework addresses and overcomes these limitations.
\item The synthetic shapes are generated from compact proper orthogonal decomposition coefficient space using Bayesian posterior sampling that relies on the No-U-Turn sampler. The estimate itself is constructed using priors fitted to real shapes. The likelihood relies on a machine-learning classifier to delineate regions of good and bad shapes in compact shape space.
\end{highlights}


\end{frontmatter}



\section{Introduction}\label{sec_introduction}


Over the last decade, tremendous progress has been made in the field of medical image analysis. Deep learning methods have become ubiquitous across a variety of medical image analysis tasks and offer state-of-the-art performance \citep{azad_advances_2024,litjens_survey_2017}. Nevertheless, their performance often depends on access to sufficiently large and well-annotated imaging datasets \citep{chen_recent_2022}. Two common bottlenecks are scarce annotation, where high-quality annotations exist only for small datasets, and weak annotation, where larger datasets are available but labels are partial or noisy. Tajbakhsh et al. (2020) offer a comprehensive review of complementary methods for addressing scarce and weak annotations \citep{tajbakhsh_embracing_2020}. One method is synthetic augmentation. In this context, generative models such as generative adversarial networks (GANs), variational autoencoders (VAE), transformation networks, and diffusion models (DM) have been successfully applied \citep{ibrahim_generative_2025, guibas_synthetic_2018, tajbakhsh_embracing_2020, Fu_2018_CVPR_Workshops, Zhao_2019_CVPR}. They are used to synthetically generate or extend existing datasets for a specific modality and anatomical structure.
However, problems remain: First, they themselves can be data-intensive, which is problematic if no existing, openly available, moderate- to large-sized dataset is available. Second, they operate on the voxel (pixel) level without guaranties on the physiological validity of anatomical structures in the generated images or masks. 

This is especially true for heart valves. Atrioventricular valves, i.e., the mitral and the tricuspid valves, are routinely imaged using three-dimensional transesophageal echocardiography (3D TEE). Despite the high volume of procedures that produce 3D TEE data, few annotated valve datasets are available because expert labeling and segmentation remain costly \citep{munafo_deep_2024}. This limits downstream tasks such as automated valve segmentation, where current models are often trained on small datasets and could benefit from larger training cohorts \citep{herz_segmentation_2021, aly2022fully, pop_image_2017, ivantsits_mv-gnn_2024}. 
But for heart valves, the object needed by many downstream tasks is not only the image itself but the segmented valve geometry. Valve shape is the link between image analysis, synthetic image-mask generation, morphological analysis, and in silico mechanics \citep{mehari_abraha_morphological_2026}. This suggests a different route to data augmentation: rather than generating images directly at the voxel level, one can generate physiologically plausible valve shapes in a compact geometric representation and then use those shapes to create masks, synthetic images, or computational models.

Making this shape-first strategy practical requires a representation that is expressive enough to capture valve anatomy, consistent enough to compare shapes across subjects, and structured enough to support generative modeling. Earlier shape representations used vectors to encode specific lengths, sizes, and angles, but these are incapable of capturing the complexities of shape \citep{lindner_automated_2017}. To better represent these complexities, many contemporary representations exist, such as: (i) point clouds \citep{zhang_survey_2025}, (ii) signed-distance functions \citep{park_deepsdf_2019}, (iii) polygonal meshes (graphs) \citep{lin_pixels_2026}, or (iv) non-uniform rational B-spline surfaces (NURBS).
After establishing a common representation and correspondence across subjects, collections of shapes can be used to build statistical shape models (SSMs) that capture the mean anatomy and its principal modes of variation \citep{rea_biomedical_2019}. They have recently been successfully applied for cardiac shapes \citep{scuoppo_generation_2025, barnet_wide_2025, amin_euclidean_2024, aljassam_machine_2024, sophocleous_analysing_2022, hoeijmakers_combining_2020}. The most common choice for forming shape models is principal component analysis (PCA) \citep{ambellan2019statistical}. Often, the first few principal components are used, and samples are generated within two or three standard deviations of the mean, which is then used as a synthetic shape generator \citep{hoeijmakers_combining_2020, mehari_abraha_morphological_2026}. This process implicitly assumes that the PCA coefficients of the shapes follow a multivariate Gaussian distribution; however, this is generally false. The sampling also ignores whether the drawn shapes are physiological, of a specific morphological type, or free of defects arising from complex representations. For example, heart valves are increasingly represented using NURBS \citep{moola_valvefit_2026, sacks2022neural}, but a PCA-based SSM commonly produces shapes with defects such as non-physiological kinks or folds, as shown in Figure \ref{fig_testproblems}(b) and (c). Clearly, there is a need for a synthetic shape generator model that can work with (i) shapes that do not follow a multivariate Gaussian distribution in PCA coefficient space (see Section 3.1 of \citep{bhalodia_deepssm_2024}), (ii) have a mechanism to reject shapes based on certain criteria, and (iii) work with contemporary shape representations. In addition to conveying shape information, the ability to draw shapes conditioned on specified criteria, such as physiological geometry or morphological type, is useful. 
In this work, we address this need by formulating synthetic heart-valve shape generation as Bayesian posterior sampling in a compact geometric coefficient space. This differs from image-level generative models, which synthesize data directly at the voxel level, and from conventional PCA-based SSMs, which sample coefficient vectors from prescribed Gaussian-like ranges without a native mechanism to enforce validity. Instead, we first represent each valve as a structured geometric object and map these shapes into a low-dimensional coefficient space. In this space, we construct a data-driven generative prior from the available real shapes and combine it with a classifier-based likelihood that favors geometrically valid, physiologically plausible reconstructions. Sampling from the resulting posterior allows us to generate synthetic valve shapes that reflect the observed shape distribution while avoiding defective or non-physiological geometries.

We demonstrate the framework for heart valves imaged with three-dimensional ultrasound, focusing on tricuspid valves from 3D TEE. The method is designed for low-data settings, where only small cohorts of expert-segmented valves are available, and we evaluate its performance starting from datasets as small as $N=10$. We first verify the approach on a controlled model problem, then validate it on parametrically generated aortic valves, and finally apply it to adult human tricuspid valves. Beyond shape generation itself, we show that the generated valves can serve as reusable geometric inputs for downstream tasks, including \textit{in silico} valve mechanics and synthetic image-mask generation. In the latter case, we close the loop by using the ray-tracing ultrasound simulation framework of \citet{spencer_fully_2026} to generate synthetic 3D ultrasound images from the synthetic valve shapes.

The remainder of the paper follows this shape-generation pipeline. Section \ref{sec_methods} first introduces the NURBS valve representation (Section \ref{sec_shaperepresentation}) and the construction of a compact PCA coefficient space (Section \ref{sec_POD}). We then review PCA-based statistical shape models as the baseline approach (Section \ref{sec_methods_ssm}) before formulating the proposed Bayesian posterior sampler (Section \ref{sec_bayesian}), including the data-driven generative prior (Section \ref{sec_generatormodel}), the classifier-based validity likelihood (Section \ref{sec:classifier}), and the NUTS sampling procedure (Section \ref{sec_hmc_nuts}). Section \ref{sec_testproblems} describes the three test problems used to evaluate the method: a low-dimensional verification problem, a parametrically generated aortic-valve validation problem, and an application to 3D TEE-derived tricuspid valves. We then compare the proposed sampler against PCA-based SSMs and generator-rejector sampling, and finally demonstrate how the generated shapes can support synthetic 3D ultrasound image-mask generation and finite-element valve simulations.

\section{Methods}\label{sec_methods}
All valve shape models in this work start from valve geometries that are represented in a common coordinate format. For image-derived valves, these geometries come from segmentations; for the aortic-valve validation problem, they are generated parametrically. We first convert each valve geometry into a common NURBS representation, which provides corresponding control-point vectors across subjects. From this shared data matrix, we construct two shape-generation models. The first is a PCA-based statistical shape model, which serves as the baseline. The second is the proposed Bayesian posterior sampler, which uses a compact coefficient space, a data-driven prior, and a classifier-based validity likelihood to generate physiologically plausible valve shapes.
\subsection{NURBS Shape Representation}\label{sec_shaperepresentation}
For each valve geometry, we use a NURBS surface representation \citep{piegl_nurbs_1997}. This representation captures the valve surface while preserving a consistent control-point structure across subjects.
Each fit starts from the same cylindrical template, whose axis is aligned with the flow through the valve and whose diameter is chosen to enclose the full valve geometry. The ValveFit framework \citep{moola_valvefit_2026} deforms this template to the segmented valve surface by updating the coordinates of the control-point mesh. All other NURBS quantities, including the order, degree, knot vectors, and control-point layout, are kept fixed. As a result, every valve is represented by a surface with the same number and ordering of control points in the circumferential and axial directions.
Thus, we express each shape as a vector of control points of the same size, $\mathbf{a}_0^{(i)} \in \mathbb{R}^{d_0}$. Here, $d_0=3C_{p_1}C_{p_2}$, where $C_{p_1}$ and $C_{p_2}$ denote the number of control points along each parametric direction; the factor of $3$ accounts for the spatial coordinates, and $(i)$ indexes the $N$ shape samples. Stacking $N$ such shapes yields the $\mathbf{A}_0$ data matrix:
\begin{equation}
    \mathbf{A}_0 = \begin{bmatrix} \mathbf{a}_0^{(1)} \\ \vdots \\ \mathbf{a}_0^{(N)} \end{bmatrix}
    \in \mathbb{R}^{N \times d_0}. \label{eq_datamatrix}
\end{equation} 

\subsection{Baseline: PCA-based Statistical Shape Models}\label{sec_methods_ssm}
We use PCA-based SSMs as the baseline approach for shape generation. Given the corresponding control-point vectors introduced in Section \ref{sec_shaperepresentation}, the baseline computes a mean shape and applies PCA to the centered deviations from this mean \citep{heimann_statistical_2009, biglino_computational_2017}. This construction has, for example, been applied to heart-valve geometries \citep{hoeijmakers_combining_2020}. Briefly, the mean shape is computed by averaging over all $N$ samples:
\begin{equation}
    \mathbf{a}_0 = \frac{1}{N}\sum_{i=1}^{N} \mathbf{a}_0^{(i)} \in \mathbb{R}^{d_0}. \label{eq_datamean}
\end{equation}
The covariance matrix $\mathbf{C} \in \mathbb{R}^{d_0 \times d_0}$ is then computed from the centered data:
\begin{equation}
   \mathbf{C} = \frac{1}{N-1}\sum_{i=1}^{N} \bigl(\mathbf{a}_0^{(i)} - \mathbf{a}_0\bigr)\bigl(\mathbf{a}_0^{(i)} - \mathbf{a}_0\bigr)^T. \label{eq_ssm_covar}
\end{equation}
The eigenvalues and eigenvectors of $\mathbf{C}$ are denoted by $\lambda_j$ and  $\boldsymbol{\phi}_j \in \mathbb{R}^{d_0}$, respectively, where index $j$ is numbered to sort eigenvalues from largest to smallest. Then any shape $\mathbf{a}_0^{(i)}$ can be approximated by a linear combination of $\eta$ shape modes:
\begin{equation}
    \hat{\mathbf{a}}_0^{(i)} = \mathbf{a}_0 + \sum_{j=1}^{\eta} \alpha_{ij}\,\boldsymbol{\phi}_j, \qquad i \in \{1, 2, \ldots, N\},
\end{equation}
where the shape coefficients $\alpha_{ij}$ are obtained by projecting the centered data onto the shape modes:
\begin{equation}
    \alpha_{ij} = \boldsymbol{\phi}_j^T \cdot \bigl(\mathbf{a}_0^{(i)} - \mathbf{a}_0\bigr).
\end{equation}
The coefficient vector $\boldsymbol{\alpha}_i = \{\alpha_{ij} : j = 1, \ldots, \eta\}$ represents the compact low-dimensional representation of shape. Now, synthetic shapes can be generated by sampling $\boldsymbol{\alpha}$ using a chosen rule. Often, PCA-based SSMs use the first few components and draw coefficients $\alpha_{ij}$ from the interval $[- p\sqrt{\lambda_j} \, , \, p\sqrt{\lambda_j}]$, where $p\in \{2,3\}$\citep{hoeijmakers_combining_2020, mehari_abraha_morphological_2026}.

\subsection{Low-dimensional Coefficient Space for the Proposed Sampler}\label{sec_POD}
The proposed sampler uses a low-dimensional representation, similar to one described in Section \ref{sec_methods_ssm}, but changes how new coefficient vectors are generated. 
Rather than sampling each mode independently from a prescribed interval, we use the coefficient space as the domain for a data-driven prior and a classifier-based validity likelihood, introduced in Section \ref{sec_bayesian}. 
This section constructs that coefficient space from the NURBS control-point matrix $\mathbf{A}_0$ defined in Eq.~\eqref{eq_datamatrix}.

\paragraph{Normalization} Before computing the coefficient space for the proposed sampler, we rescale the shape-vector matrix coordinate-wise. This preprocessing differs from the PCA-based SSM baseline in Section \ref{sec_methods_ssm}, where PCA is applied directly to mean-centered shape vectors in the original coordinate scale. This is done to avoid issues in the sampling procedure arising from differences in relative scales. Specifically, we apply min-max scaling to affinely transform each coordinate of the shape vectors to $[0,1]$,
\begin{align}
    (\mathbf{a}_0^{min})_j = \min_{i}\, (\mathbf{a}_0^{(i)})_j,
    &\quad
    j\in[1,d_0]\\
    (\mathbf{a}_0^{max})_j = \max_{i}\, (\mathbf{a}_0^{(i)})_j,
    &\quad 
    j\in[1,d_0]\\
    \boldsymbol{\Delta}_0 = \mathbf{a}_0^{max} - \mathbf{a}_0^{min} + \varepsilon,
    &\quad
    \left\{\mathbf{a}_0^{min}, \mathbf{a}_0^{max}, \boldsymbol{\Delta}_0 \right\} \in \mathbb{R}^{1 \times d_0},
\end{align}
where $\varepsilon>0$ is to prevent division by zero. The min-max scaled matrix is,
\begin{equation}
    \widehat{\mathbf{A}}_0 = \bigl(\mathbf{A}_0 - \mathbf{1}_N \mathbf{a}_0^{min}\bigr) \oslash \bigl(\mathbf{1}_N \boldsymbol{\Delta}_0\bigr)
    \in [0,1]^{N\times d_0},
\end{equation}
where $\oslash$ is element-wise (Hadamard) division and $\mathbf{1}_N \in \mathbb{R}^{N\times 1}$ is a column vector of ones. The scaled data are then centered by subtracting their column-wise mean,
\begin{equation}
    \hat{\boldsymbol{\mu}}_0 = \frac{1}{N}\,\mathbf{1}_N^\top \widehat{\mathbf{A}}_0 \in \mathbb{R}^{1\times d_0},
    \qquad
    \widetilde{\mathbf{A}}_0 = \widehat{\mathbf{A}}_0 - \mathbf{1}_N\hat{\boldsymbol{\mu}}_0. \in \mathbb{R}^{N \times d_0}
\end{equation}
Note that we store all statistics for future transformations and inversion. 

\paragraph{Covariance and POD}
The covariance matrix of the normalized, centered data is,
\begin{equation}
    \mathbf{C}_0 = \frac{1}{N-1}\,\widetilde{\mathbf{A}}_0^\top\widetilde{\mathbf{A}}_0 \in \mathbb{R}^{d_0 \times d_0}.
\end{equation}
Its eigendecomposition yields the eigenmodes (or POD-modes) $\boldsymbol{\phi}_i \in \mathbb{R}^{d_0}$ and eigenvalues $\lambda_i$, again sorted in descending order,
\begin{equation}
    \mathbf{C}_0\,\boldsymbol{\phi}_i = \lambda_i\,\boldsymbol{\phi}_i, \qquad \lambda_1 \geq \lambda_2 \geq \cdots \geq \lambda_{d_0} \geq 0.
\end{equation}
Assembling the modes column-wise into $\boldsymbol{\Phi}_0 = [\boldsymbol{\phi}_1 \mid \cdots \mid \boldsymbol{\phi}_{d_0}] \in \mathbb{R}^{d_0 \times d_0}$, we can project each shape in the dataset along these modes to obtain the POD coefficients:
\begin{equation}
    \widetilde{\mathbf{A}}_1 = \widetilde{\mathbf{A}}_0\,\boldsymbol{\Phi}_0 \in \mathbb{R}^{N \times d_0}.\label{eq_podcoefficients}
\end{equation}
Accordingly, each shape can be expressed as a linear combination of the POD coefficients and the corresponding POD modes. We define the cumulative fractional energy of the first $m$ modes as:
\begin{equation}
    \mathcal{E}(m) = {\displaystyle\sum_{i=1}^{m}\lambda_i}\bigg/{\displaystyle\sum_{i=1}^{d_0}\lambda_i}.\label{eq_fracenergy}
\end{equation}
We truncated the POD coefficients to the first $k$ modes that exceed a certain energy threshold $E_T$, which we set to 95\%. Note that the number of non-zero eigenvalues is restricted by the rank of the covariance matrix $\mathbf{C}_0$. For real-valued $\widetilde{\mathbf{A}}_0$ for which the mean row has been subtracted from the matrix, the rank can at most be $\min\left\{N-1, d_0\right\}$. Thus, the retained dimension is
\begin{equation}
    d_1 = \min\bigl\{m : \mathcal{E}(m) \geq E_{\text{T}}, N-1,d_0\bigr\}. \label{eq_d1_select}
\end{equation}
The truncated coefficient matrix is then obtained as
\begin{equation}
    \mathbf{A}_1 =
    \begin{bmatrix}
        \mathbf{a}_1^{(1)} \\
        \vdots \\
        \mathbf{a}_1^{(N)}
    \end{bmatrix}
    =
    (\widetilde{A}_{1,ij})_{{i\in[1,N]},{j\in[1,d_1]}}
    \in\mathbb{R}^{N \times d_1},
    \label{eq_truncatedpod1}
\end{equation}
where $\mathbf{a}_1^{(i)} \in \mathbb{R}^{d_1}$ denotes the truncated POD coefficient vector of the $i$th real shape. All subsequent generative modeling steps are performed in this truncated coefficient space.  Thus, after the representation and projection steps above, each real valve is represented by a point in $\mathbb{R}^{d_1}$, and generating a new valve amounts to sampling a new point in this space that can be mapped back to a NURBS surface.

\subsection{Bayesian Posterior Sampling}\label{sec_bayesian}
The remaining task is to define a probability distribution over the truncated POD-coefficient space. We denote a single synthetic coefficient vector by $\mathbf{a}_1^{(s)} \in \mathbb{R}^{d_1}$ and a collection of $N_s$ synthetic samples by $\mathbf{A}_1^{\mathrm{syn}} \in \mathbb{R}^{N_s \times d_1}$. Each sampled vector $\mathbf{a}_1^{(s)}$ can be mapped back to a NURBS surface by reconstructing the scaled and centered control-point vector and then reversing the centering and scaling transformations from Section \ref{sec_POD}.

The real coefficient matrix $\mathbf{A}_1$ provides samples from the unknown distribution of physiologically observed valve shapes. Because this distribution may be multimodal and non-Gaussian, we first fit a data-driven generator $\mathcal{G}$ that defines a prior distribution over the coefficient space. However, samples from this prior may still correspond to defective or non-physiological reconstructed shapes, such as shapes with kinks, folds, or self-intersections. We therefore introduce a classifier $\mathfrak{C}$ that estimates whether a sampled coefficient vector corresponds to a \textit{good} (physical) or \textit{bad} (non-physical) shape. The definition of these classes and the construction of the classifier are described in Section \ref{sec:classifier}.

We formulate synthetic shape generation as sampling from the distribution of real valve shapes conditioned on geometric validity. In this formulation, the generator $\mathcal{G}$ defines a prior over the truncated POD-coefficient space, while the classifier $\mathfrak{C}$ provides a likelihood that favors coefficient vectors whose reconstructed shapes are classified as \textit{good}. By Bayes' theorem,
\begin{equation}
    \pi(a \mid b)
    = \frac{\pi(b \mid a)\,\pi(a)}{\pi(b)}
    \propto \pi(b \mid a)\,\pi(a),
\end{equation}
where $\pi(a)$ is the prior, $\pi(b \mid a)$ is the likelihood, and $\pi(b)$ is the evidence. In the present setting, this gives
\begin{align}
    \pi[\text{shape} \mid \text{good}]
    &\propto
    \pi[\text{good} \mid \text{shape}]
    \, \pi[\text{shape}].
    \label{eq_bayes_our_language}
\end{align}

More specifically, let $c=0$ and $c=1$ denote \textit{bad} and \textit{good} shapes, respectively. For a synthetic coefficient vector $\mathbf{a}_1^{(s)} \in \mathbb{R}^{d_1}$, the target posterior is
\begin{align}
    \pi\!\left(\mathbf{a}_1^{(s)} \mid c(\mathbf{a}_1^{(s)})=1\right)
    &\propto
    \underbrace{
    \pi_{\mathfrak{C}}\!\left(c(\mathbf{a}_1^{(s)})=1 \mid \mathbf{a}_1^{(s)}\right)
    }_{\text{validity likelihood}}
    \,
    \underbrace{
    \pi_{\mathcal{G}}\!\left(\mathbf{a}_1^{(s)}\right)
    }_{\text{generative prior}} .
    \label{eq_bayes_our}
\end{align}

The posterior in Eq.~\eqref{eq_bayes_our} is known only up to a proportionality constant, since the evidence is not computed explicitly. We therefore draw samples from this target distribution using Hamiltonian Monte Carlo, specifically the No-U-Turn sampler described next.

\begin{figure*}[tbp]  
	\centering
	\includegraphics[width=\textwidth]{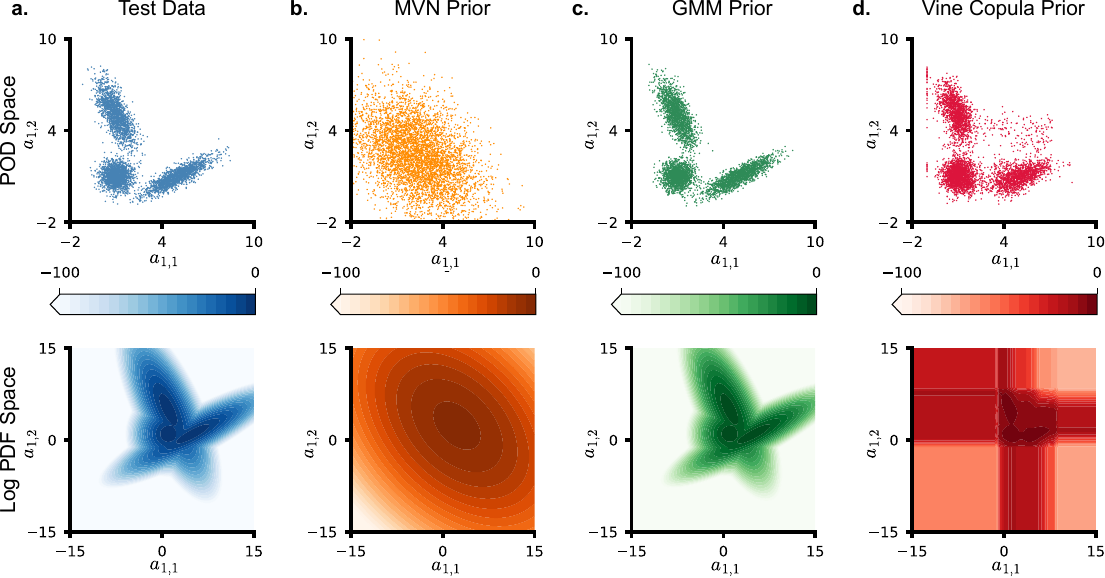}
	\caption{Various priors fit to the generated test data, where the top row shows points in $\mathbf{A}_1$ space, and the bottom row shows the log-PDF space. \textbf{(a)} The ground truth test data. \textbf{(b)} Multivariate normal prior fit to the ground truth as is traditionally used in SSMs. \textbf{(c)} Gaussian mixture model prior fit to the ground truth as chosen in the current work. \textbf{(d)} Vine copula prior fit to the test data, which is not used due to pathological behavior around high marginals in the log-probability space.}\label{fig_senstivity_prior}
\end{figure*}

\subsubsection{Hamiltonian Monte Carlo (HMC) and No-U-Turn Sampler (NUTS)}\label{sec_hmc_nuts}

\paragraph{HMC} Instead of finding a model for the posterior, using, for example, variational inference (VI) \citep{blei_variational_2017}, we directly sampled from it. A common posterior-sampling method is Markov chain Monte Carlo (MCMC)  \citep{van_ravenzwaaij_simple_2018}. A more efficient version of MCMC is Hamiltonian Monte Carlo (HMC) \citep{betancourt_conceptual_2018}. In the current work, we use the No-U-Turn sampler (NUTS) \citep{hoffman2014no}, an adaptive variant of HMC. Briefly, the HMC method generates samples from the `target distribution' $\pi(\mathbf{q})$. It does so by constructing a joint probability distribution called the `canonical distribution' $\pi(\mathbf{q},\mathbf{p})$, where $\mathbf{p}$ is the momentum and $\mathbf{q}$ is the position. HMC proposes moves in the joint phase space of $(\mathbf{p},\mathbf{q})$:
\begin{equation}
    \pi(\mathbf{q},\,\mathbf{p}) = \pi(\mathbf{p} \mid \mathbf{q})\;\pi(\mathbf{q})\, .
\end{equation}
The Hamiltonian is defined as the negative log-joint probability:
\begin{align}
    H(\mathbf{q},\,\mathbf{p})
    &:= -\log\,\pi(\mathbf{q},\,\mathbf{p})
    \notag\\
    &= \underbrace{-\log\,\pi(\mathbf{p} \mid \mathbf{q})}_{K(\mathbf{p},\,\mathbf{q})\;\text{(kinetic energy)}}
       +
       \underbrace{\bigl(-\log\,\pi(\mathbf{q})\bigr)}_{V(\mathbf{q})\;\text{(potential energy)}} \, .
\end{align}
where $H$ is the Hamiltonian, $K$ is the kinetic energy model, and $V$ is the potential energy. The latter $\pi(\mathbf{q})$ represents our target posterior distribution from Equation \ref{eq_bayes_our}. Thus,
\begin{align}
    V(\mathbf{q}) &= -\log\,\pi\bigl[\mathbf{q} \mid c(\mathbf{q})=1\bigr] \notag\\
         &= -\log\Bigl(\pi\bigl[c(\mathbf{q})=1 \mid \mathbf{q}\bigr]\cdot\pi(\mathbf{q})\Bigr) -\log C\notag\\
         &= \underbrace{-\log\,\pi\bigl[c(\mathbf{q})=1\mid \mathbf{q}\bigr]}_{\text{likelihood}-\mathfrak{C}}
            \;+\;
            \underbrace{-\log\,\pi(\mathbf{q})}_{\text{prior}-\mathcal{G}} -\log C,\label{eq_potential_our}
\end{align}
where $\log C$ is an additive constant. A practical HMC uses an outer Metropolis-Hastings algorithm \citep{hitchcock_history_2003} and inner Hamiltonian trajectories simulated with a symplectic integrator, typically a leapfrog scheme (see details in \citep{betancourt_conceptual_2018}).
These integrators require $\partial V/\partial \mathbf{q} $ to be computable. Any model used to compute potential energy, including the chosen generator $\mathcal{G}$ and classifier $\mathfrak{C}$, must provide differentiable negative log-probability terms with respect to the input coefficient vector. We obtain these derivatives through automatic differentiation, which motivates our use of PyTorch \citep{paszke2017automatic} and JAX \citep{jax2018github} for implementing the generator $\mathcal{G}$ and the classifier $\mathfrak{C}$.

\paragraph{NUTS} Two key challenges in the use of HMC are setting the step size of the symplectic integrators and the length of the Hamiltonian trajectories. The No-U-Turn sampler (NUTS) is an adaptation of HMC that addresses both of these concerns \citep{hoffman2014no}. We implement NUTS using the PyMC Bayesian modeling library \citep{pymc2023}. Specifically, we combine the generator described in Section \ref{sec_generatormodel} and the classifier described in Section \ref{sec:classifier} as the prior and likelihood terms of the posterior in Eq.~\eqref{eq_bayes_our}. We run multiple independent chains of the NUTS sampling procedure. We discard the initial warm-up samples and pool the remaining samples across chains. Further implementation details are provided in \ref{sec_appendix_nuts}.

\subsubsection{Generative Prior over the Truncated Coefficient Space}\label{sec_generatormodel}
We use the truncated POD coefficient matrix $\mathbf{A}_1$ (Equation \ref{eq_truncatedpod1}) from real shapes to fit a model $\mathcal{G}$ which we then use as a generator. In general, the coefficient vectors ${\mathbf{a}_1^{(i)}}$ have a distribution with multiple modes and might not necessarily conform to any standard distribution. The sampling procedure traditionally used in SSM, as described in Section \ref{sec_methods_ssm}, relies on a multivariate normal (MVN) distribution to fit points that lie in the truncated space of the first few coefficients. This construction imposes unimodality and ellipsoidal structure on the joint distribution. To relax these constraints, we experimented with two specific priors: (i) a Gaussian mixture model (GMM), and (ii) a Vine copula \cite{torchvinecopulib}. Both priors are more flexible than a single MVN for multimodal or non-ellipsoidal coefficient distributions. See the top row of Figure \ref{fig_senstivity_prior} to visualize how each prior does on generic test data. The test data is drawn from a Gaussian mixture model with multiple modes, each with a distinct ellipsoidal covariance that need not be aligned with the coordinate axes. Note that for our GMM prior, we did not specify the number of mixtures; instead, we set an upper limit of $10$. The exact number of modes chosen is guided by the available data and the default selection criteria in the Scikit-learn library \citep{scikit-learn}. Our motivation for considering a vine copula was its generality in capturing multivariate distributions \citep{torchvinecopulib} and its ability to decouple marginal behavior from the dependence structure. 

\begin{figure*}[htbp]  
	\centering
	\includegraphics[width=\textwidth]{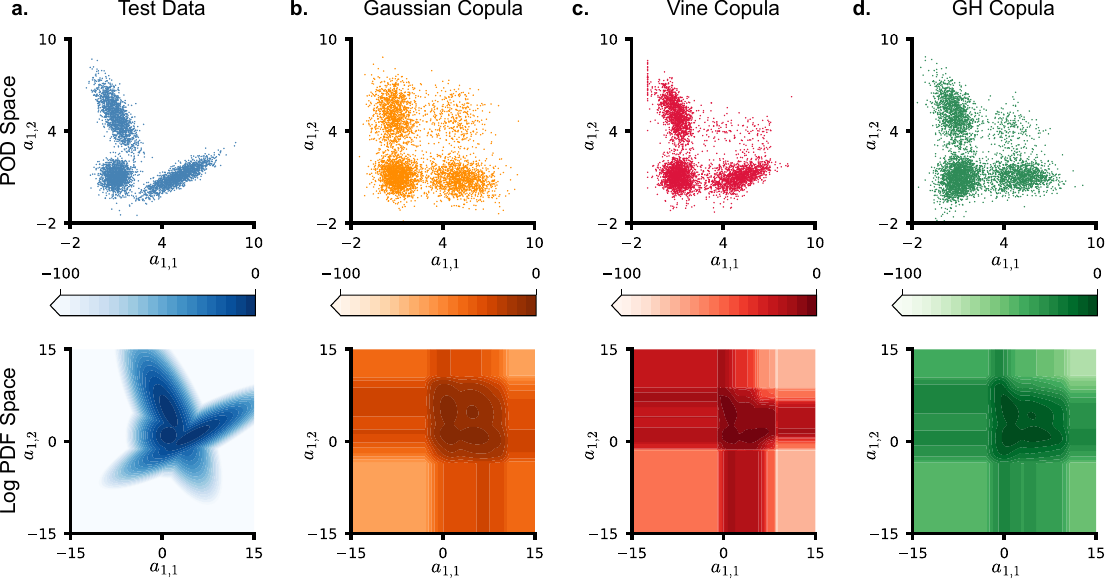}
	\caption{Various copula prior fits to the generated test data, where the top row shows points in $\mathbf{A}_1$ space, and the bottom row shows the log-PDF space. \textbf{(a)} The ground truth test data, \textbf{(b)} the Gaussian copula prior with custom implementation, \textbf{(c)} Vine copula prior implemented using \texttt{torchvinecopulib} \citep{torchvinecopulib}, and \textbf{(d)} generalized hyperbolic copula prior implemented using \texttt{CopulAX} \citep{tfm000_copulax}. Note the pathological behavior in log-PDF space that attracts NUTS trajectories along bands of high marginals.}\label{fig_senstivity_copula}
\end{figure*}

The second row of Figure \ref{fig_senstivity_prior} shows the log-probability space returned by the fit prior and comparison to that of the test data. As expected, the MVN prior used in typical SSMs performs worse than the GMM and Vine copula priors in the neighborhood of the test data. More precisely, inside a constant Mahalanobis distance boundary that contains the data, both the GMM and the copula prior do better. However, outside of this boundary, we note that both MVN and GMM prior decay rapidly, while the copula exhibits bands that extend along coordinate axes at regions of high marginals. We experimented with all copulas available in both SklarPy \citep{sklarpy} and CopulAX \citep{tfm000_copulax}, a subset of which are shown in Figure \ref{fig_senstivity_copula}. We note that all copulas exhibited the same pathological behavior of bands of high log-probability extending far outside the neighborhood of the dataset used to fit them. This artifact proved problematic, as NUTS trajectories often got stuck while traveling along these bands. Thus, even though copulas offer a convenient way to capture multivariate distributions, we settled for using a Gaussian mixture model with a data-driven number of modes.

\subsubsection{Classifier}\label{sec:classifier}
A key requirement of our method is the ability to assess whether a sampled coefficient vector reconstructs to a valid or defective shape. We encode this using a classifier $\mathfrak{C}$ that estimates a class label $c$ for each sampled coefficient vector $\mathbf{a}_1^{(s)}$, or for features computed from its reconstructed shape. To train this classifier, we generate candidate samples with the generator $\mathcal{G}$ and assign labels to their reconstructed shapes. In the model problem, labels are assigned programmatically using the prescribed decision boundary. For the valve problems, the generated samples are manually labeled through visual inspection of the reconstructed meshes. In principle, the same labeling step could also be performed by clinicians or domain experts when the desired condition is clinical plausibility, morphology, or another expert-defined criterion.
Then, using the combined data of real shapes, all of which belong to $c=1$, and the labeled shapes, which contain both classes, we train the classifier. The classifier $\mathfrak{C}$ is a multilayer perceptron (MLP) with fully connected layers of the form $\{\text{Linear} \to \text{ReLU} \to \text{Dropout}\}$, repeated for each hidden layer, followed by a final linear layer producing two logits. Class probabilities are obtained via softmax, and $\log\pi[c=1 \mid \mathbf{a}_1^{(s)}]$ serves as the likelihood term in the potential energy (Equation \ref{eq_potential_our}). To handle label imbalance, cross-entropy loss is weighted by the inverse class frequency. All inputs are Z-score normalized before the forward pass. We use the same MLP architecture for all three test problems in this work; however, the input features and the spaces from which the features are constructed differ. Because the three test problems differ in how validity is defined, we use two classifier input spaces.

\begin{figure*}[htbp]  
	\centering
	\includegraphics[width=\textwidth]{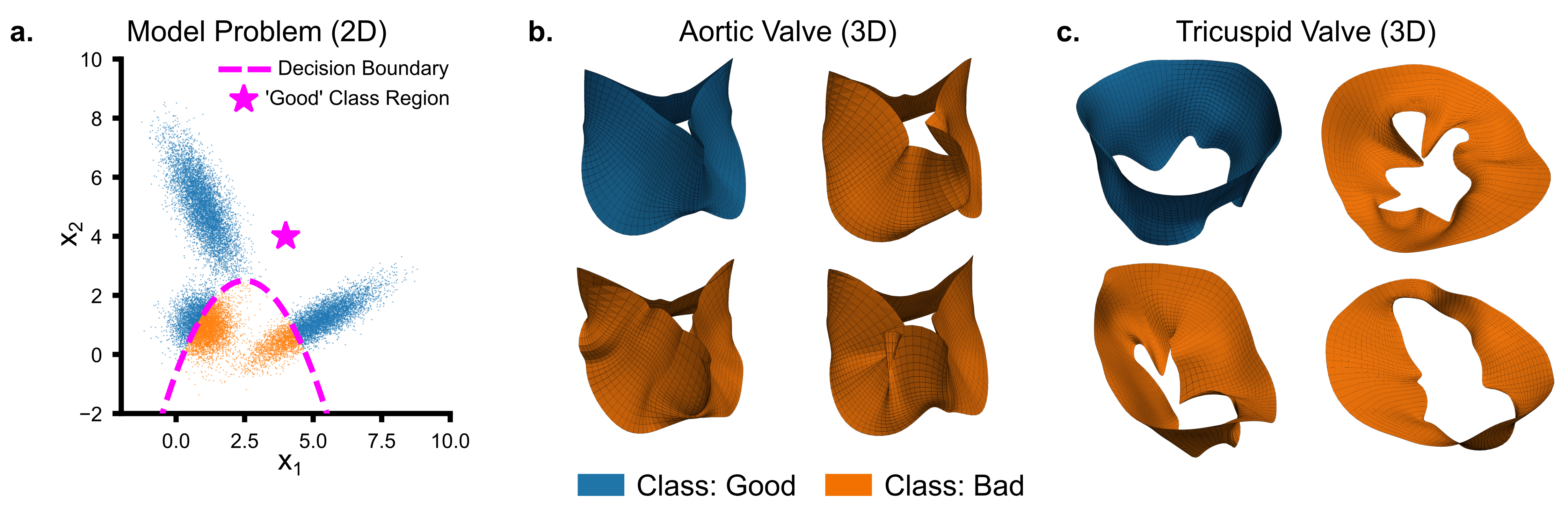}
	\caption{Various problems tested using the current framework. \textbf{(a)} Snapshot space of the two-dimensional model problem with a decision boundary that distinguishes the good and bad class labels. \textbf{(b)} Generated aortic valve shapes for both classes. \textbf{(c)} Generated tricuspid valve shapes for both classes.}\label{fig_testproblems}
\end{figure*}

\paragraph{Valve problem classifier} For the aortic and tricuspid valve cases (Sections \ref{sec_testproblem_aortic} and \ref{sec_testproblems_tricuspid}), $\mathfrak{C}$ operates in an eight-dimensional geometric quality feature space rather than directly on the truncated POD coefficient vector $\mathbf{A}_1$. A feature vector
$\mathbf{e} \in \mathbb{R}^8$ is computed from the NURBS surface reconstructed by inverting the transformation ($\mathbf{A}_1 \to \mathbf{A}_0$).
\begin{align}
    \mathbf{e} &:= \left[R_{tpe}^{max}, R_{ortho}, R_{norm}, \theta_{\mathbf{t}_1\cdot \mathbf{t}_2}^{max},
                     \mathcal{L}, C_{\| \mathbf{t}_1 \times \mathbf{t}_2 \|}, C_{sym}, P^{\hat{\mathbf{n}}}_{0.05}\right]^T \label{eq_class_ip_valve}
\end{align}
Of these, the first three features are based directly on the regularization terms used in ValveFit \citep{moola_valvefit_2026}. The remaining five are added in the current work. These quantities capture local and global indicators of geometric quality. Specifically, they enable the classifier to detect self-intersection, local non-orthogonality, distortion, kinks, and, where applicable, symmetry. 
Some of these defects are shown in Figure \ref{fig_testproblems} (b) and (c). 
All quantities below are evaluated on a sampled surface mesh, referred to here as the evaluation mesh, not on the NURBS control-point mesh used to define the surface in Section \ref{sec_shaperepresentation}. An element denotes one surface patch bounded by adjacent parameter values in the two parametric directions. We define each feature below.

\subparagraph{Maximum tangent point energy $(R_{tpe}^{max})$}
This feature is defined as the maximum over all pairs of distinct points on the surface. Given the surface points $\mathbf{s}_j$ and $\mathbf{s}_k$,
\begin{equation}
    R_{tpe}^{max} = \max_{j \neq k}\; \frac{\bigl|\mathbf{n}_k \cdot (\mathbf{s}_j - \mathbf{s}_k)\bigr|^\alpha}{\|\mathbf{s}_j - \mathbf{s}_k\|^{2\alpha}},
\end{equation}
where $\alpha = 2$ by default and $\mathbf{n}_k = \mathbf{t}_1^{(k)} \times \mathbf{t}_2^{(k)}$ is the surface normal at point $k$. This term is high if there are global self-intersections in the surface. While used as a regularization term to reduce self-intersection in ValveFit, we use it here as an indicator of self-intersections. For more details, see Figure 9 of \citep{moola_valvefit_2026}.

\subparagraph{Tangent orthogonality energy $(R_{ortho})$}
The tangent orthogonality energy is the mean absolute inner product between the unit tangents $\hat{\mathbf{t}}_1$ and $\hat{\mathbf{t}}_2$, averaged over all $M$ points on the surface:
\begin{equation}
    R_{ortho} = \frac{1}{M}\sum_{i=1}^{M} \bigl|\hat{\mathbf{t}}_1^{(i)} \cdot \hat{\mathbf{t}}_2^{(i)}\bigr|.
\end{equation}
A value of zero indicates perfectly orthogonal parametric directions everywhere; values near one signal extreme element distortion. 

\subparagraph{Normal deviation energy $(R_{norm})$} The normals $\mathbf{n}^{(i)} \in \mathbb{R}^3$ at each point are scaled to $[0,1]$ using a single scalar minimum and maximum taken over all vectors and components:
\begin{align}
    \tilde{\mathbf{n}}^{(i)} &= \frac{\mathbf{n}^{(i)} - n_{\min}}{n_{\max} - n_{\min}},\\
    n_{\min} = \min_{i,\,k}\, n_k^{(i)},
    &\quad
    n_{\max} = \max_{i,\,k}\, n_k^{(i)}, 
\end{align}
where $n_k^{(i)}$ is the $k$-th Cartesian component of $\mathbf{n}^{(i)}$. Then $R_{norm}$ is computed as:
\begin{align}
    R_{norm} &= \max_i \Bigl|{n}_z^{(i)} - \frac{1}{M}\sum_{j=1}^{M}\tilde{n}_z^{(j)}\Bigr|.
\end{align}
$R_{norm}$ helps indicate abrupt changes in surface normal. 

\subparagraph{Max.\ tangent inner product $(\theta_{\mathbf{t}_1\cdot \mathbf{t}_2}^{max})$}
While $R_{ortho}$ averages non-orthogonality globally, this feature captures the single worst-case location:
\begin{equation}
    \theta_{\mathbf{t}_1\cdot \mathbf{t}_2}^{max} = \max_i \bigl|\hat{\mathbf{t}}_1^{(i)} \cdot \hat{\mathbf{t}}_2^{(i)}\bigr|.
\end{equation}
This indicates if a severe local non-orthogonality exists. 

\subparagraph{Log max.\ element aspect ratio $(\mathcal{L})$} The aspect ratio at point $i$ is  $\mathcal{A}_i$,
\begin{align}
    \mathcal{A}_i = \max\left(\frac{\|\mathbf{t}_1^{(i)}\|}{\|\mathbf{t}_2^{(i)}\|},\frac{\|\mathbf{t}_2^{(i)}\|}{\|\mathbf{t}_1^{(i)}\|}\right)
\end{align}
where $\mathbf{t}_1^{(i)}$ and $\mathbf{t}_2^{(i)}$ are tangent vectors in two parametric directions. We apply log-compression on the worst aspect ratio in the entire surface,
\begin{equation}
    \mathcal{L} = \log\!\left(1 + \max_i \mathcal{A}_i\right).
\end{equation}
Log-compression ensures that very high aspect ratios reflect small changes in $\mathcal{L}$. 

\subparagraph{Coefficient of variation of element areas $(C_{\|\mathbf{t}_1\times \mathbf{t}_2\|})$} The area of each element is computed as $A^{(i)} = \|\mathbf{t}_1^{(i)} \times \mathbf{t}_2^{(i)}\|$. We obtain the coefficient of variation of this area across all elements. Once again, log-compression was applied,
\begin{align}
    \mu_A = \frac{1}{M}\sum_i A^{(i)},
    &\quad
    \sigma_A = \sqrt{\frac{1}{M}\sum_i \bigl(A^{(i)}-\mu_A\bigr)^2}, \\
    C_{\|\mathbf{t}_1\times \mathbf{t}_2\|} &= \log\!\left(\frac{\sigma_A}{\mu_A}\right) \, .
\end{align}

\subparagraph{Coefficient of variation of per-leaflet areas $(C_{sym})$}
An aortic valve comprises three leaflets and is roughly three-fold rotationally symmetric. We expect the surface areas when three-fold equipartitioned in the circumferential direction to be similar. $C_{sym}$ captures the coefficient of variation in those areas. For the aortic valve, this feature could distinguish samples labeled as bad due to gross asymmetry. For tricuspid valves, this coefficient is not relevant.  
\begin{align}
    \mu_{A_\ell} = \frac{1}{L}\sum_{\ell=1}^{L}A_\ell,
    &\quad
    \sigma_{A_\ell} = \sqrt{\frac{1}{L}\sum_{\ell=1}^{L}\bigl(A_\ell - \mu_{A_\ell}\bigr)^2}, \\
    C_{sym} &= \frac{\sigma_{A_\ell}}{\mu_{A_\ell}},
\end{align}
where $L=3$, and $A_\ell$ is the area of the respective partition.  

\subparagraph{5th-percentile cosine of adjacent face normals $(P^{\hat{\mathbf{n}}}_{0.05})$}

For all the points on the surface, the unit normals $\hat{\mathbf{n}}^{(i)}$ are obtained. Then we compute the inner product of each normal with the normal at the point adjacent to it $\hat{\mathbf{n}}^{(i+1)}$. The threshold that marks the lowest 5th percentile of values is stored,
\begin{equation}
    P^{\hat{\mathbf{n}}}_{0.05} = \text{Lowest 5}{\%}\text{ threshold}\!\bigl(\{\hat{\mathbf{n}}^{(i+1)} \cdot \hat{\mathbf{n}}^{(i)}\}\bigr).
\end{equation}
Low values of this feature indicate sharp folds.

We add five new features on top of the JAX-capable methods provided by ValveFit \citep{moola_valvefit_2026, jax2018github}.

\paragraph{Model problem} For the two-dimensional model problem (Section \ref{sec_testproblems_model}), $\mathfrak{C}$ operates directly in the truncated POD-coefficient space $\mathbf{A}_1 \in \mathbb{R}^{N \times d_1}$. The acceptable region is defined by a prescribed decision boundary in the original snapshot space.
Each sample is labeled according to which side of this boundary it lies on, with $c=1$ denoting good samples and $c=0$ denoting bad samples.
These labels are then assigned to the corresponding projected coefficient vectors in $\mathbf{A}_1$.
The classifier thus learns the boundary directly in the low-dimensional $\mathbf{A}_1$ space, with $d_1$ input features.

\paragraph{Implementation}

Training data are assembled by combining real shapes (all labeled as good, $c=1$) with a set of synthetically generated shapes that are manually labeled for the valve problem via visual inspection of the reconstructed meshes and programmatically labeled for the model problem. All inputs to the classifier are Z-score-normalized, and normalization statistics are stored for future use by the model. We monitored per-class accuracies and other metrics on both the training and validation sets during classifier training to avoid overfitting. The best trained models were used as classifiers. More details are provided in \ref{sec_appendix_classifier}

\subsection{The Three Models}\label{sec_threemethods}

In the results section, we present findings on the application of the described prevalent PCA-based SSMs (Section \ref{sec_methods_ssm}) and the proposed NUTS (with Generator $\mathcal{G}$-Classifier $\mathfrak{C}$) method. Additionally, given any classifier $\mathfrak{C}$, we can use it to reject bad shapes generated using the generator until we find good shapes, a model we term as `Generator-Rejector (GR)'. Henceforthrefer to these methods as \textbf{SSM}, \textbf{NUTS}, and \textbf{GR} methods, respectively. We first applied all three methods to a verification and a validation problem in Section \ref{sec_testproblems}. Finally, we applied the proposed NUTS for human tricuspid valve shape generation.

\section{Experiments}\label{sec_expt}

\subsection{Test Problems}\label{sec_testproblems}
We evaluate the proposed framework on three test problems of increasing anatomical realism. The first is a two-dimensional model problem in $\mathbb{R}^2$, which provides a known data distribution and a prescribed validity boundary. This problem is used to verify, visually and quantitatively, whether the methods capture multimodality while restricting samples to the good region. The second is an \textit{in silico} aortic-valve problem, in which valve shapes are generated parametrically and then represented with the same NURBS/POD pipeline used elsewhere in the paper. This problem serves as a controlled validation case on three-dimensional valve-like geometries. The third is the target application: adult human tricuspid valves segmented from 3D TEE images. Thus, the model problem serves as verification, the aortic-valve problem as controlled validation, and the tricuspid-valve problem as application of the method \citep{anderson_verification_2007, aiaa2022}. All three problems are shown in Figure~\ref{fig_testproblems}.

\subsubsection{Verification: Model Problem}\label{sec_testproblems_model}
We construct a two-dimensional verification problem by sampling points $\{\mathbf{a}_0^{(i)}\}$ in $\mathbb{R}^2$ from a Gaussian mixture model with three components. Each component has an ellipsoidal covariance matrix whose principal axes are not aligned with the coordinate axes. In this problem, the sampled points play the role of shape vectors, and $\mathbb{R}^2$ is treated as the data, or snapshot, space defined in Eq.~\eqref{eq_datamatrix}.

We then prescribe a nonlinear decision boundary that separates $\mathbb{R}^2$ into good and bad regions. Samples on one side of this boundary are assigned the good class label $(c=1)$, while the remaining samples are assigned the bad class label $(c=0)$. The boundary and the corresponding good region are shown in Figure~\ref{fig_testproblems}(a). The classifier $\mathfrak{C}$ is trained from these labels and approximates this known decision boundary in the projected coefficient space. This setup provides a controlled test of whether the sampling method can recover a multimodal distribution while respecting a prescribed validity constraint.

\subsubsection{Validation: Aortic Valve}\label{sec_testproblem_aortic}

The ground-truth shapes of the aortic valve are generated parametrically and used in lieu of segmentation from medical imaging data. We use this problem to validate our framework before applying it to image-derived tricuspid valves. Each shape is defined by four parameters representing specific dimensions of the valve \citep{xu_framework_2018}. These parameters are drawn independently from corresponding univariate Gaussian distributions. The means are set to physiologically meaningful values, and a common standard deviation is chosen via visual inspection of the resulting generated shapes. We then use ValveFit to fit NURBS surfaces to these parametrically generated shapes to obtain our desired shape representations, as detailed in Section \ref{sec_shaperepresentation}. Recall that the degrees, knot vectors, and number of control points are fixed, and only control point coordinates are updated during the fitting process. The dimensionality of the control point mesh is 594, where $(C_{p_1}, C_{p_2}, 3)\equiv(33,6,3)$. A typical good surface fit is shown in blue in Figure \ref{fig_testproblems}(b), note the three-fold rotational symmetry of the shape. We additionally show in Figure \ref{fig_testproblems}(b) defective shapes generated using generator $\mathcal{G}$ (Section \ref{sec_generatormodel}). In these \textit{bad} $(c=0)$ examples, defects such as kinks, folds, self-intersection, and asymmetry that motivate the use of classifier $\mathfrak{C}$ (Section \ref{sec:classifier}) are seen clearly. 

\subsubsection{Application: Tricuspid valve}\label{sec_testproblems_tricuspid}
The previous two problems serve to verify and validate the proposed method. The current problem concerns our intended application to the tricuspid valve. 
\paragraph{Dataset} We obtained a dataset of anonymized 3D TEE images of adult human tricuspid valves taken preoperatively at the Asklepios Hospital, Harburg (Germany) and the University Hospital Schleswig-Holstein (Germany). These data are used in accordance with a data agreement with these institutions and UT Austin. All data collection complied with the local institutional regulations of the respective hospitals. No other variables or imaging protocol were controlled when collecting these images. While the patients presented with a functional tricuspid valve, they may have had other heart conditions or diseases. No formal analysis of potential systematic bias in the dataset was performed in this study. We built a dataset consisting of ten (N=10) such ultrasound images and are continuing to expand the dataset size. These ultrasound images were first converted to three-dimensional Cartesian DICOM. The data were processed using the open-source 3D Slicer program \citep{kapur_increasing_2016, fedorov20123d, slicer2026}. There, we first identified the end-systolic and end-diastolic frames within the cardiac cycle. Next, the tricuspid valve annulus and leaflets were segmented manually using the SlicerHeart extension \citep{Lasso2022}. Five such segmentations are shown in Figure \ref{fig_tv_realsynthetic} (a) for the end-diastolic configuration. 

\paragraph{Using the dataset} We use the NURBS shape representation (Section \ref{sec_shaperepresentation}) and fit the shapes to the segmentations using the same process as used earlier for aortic valves (Section \ref{sec_testproblem_aortic}). Once again, the flattened control point coordinates ${\mathbf{a}_0^{(i)}}$ have dimension 594 and are used to construct the data (snapshot) matrix (Equation \ref{eq_datamatrix}). A generator $\mathcal{G}$ is fit to the truncated POD-coefficient space of the shapes (Sections \ref{sec_POD} and \ref{sec_generatormodel}). Examples of the shapes generated by the fitted generator $\mathcal{G}$ are shown in Figure \ref{fig_testproblems}(c). The figures shows examples of both \textit{good} shapes $(c=1)$ and \textit{bad} shapes $(c=0)$.

\subsection{Subsets}\label{sec_expt_subsets}

The performance of each shape-generation method depends on the number of available training shapes. We therefore evaluate the PCA-based SSM, the generator-rejector (GR), and the proposed NUTS sampler on subsets of varying sizes.

For the first two problems (Sections \ref{sec_testproblems_model} and \ref{sec_testproblem_aortic}), the data are generated \textit{in silico}, allowing us to construct multiple subsets with varying sample sizes. We first draw a large parent set of shape samples and then construct progressively smaller subsets from it. This design allows us to quantify how each method performs as the available training data decreases; the corresponding results are reported in Section~\ref{sec_results}.
The subset construction is more involved for the aortic-valve problem than for the model problem. In the model problem, class labels are assigned algorithmically from the prescribed decision boundary. In the aortic valve problem, the generated shapes must be manually labeled via visual inspection to train the classifier $\mathfrak{C}$. We therefore construct fewer subsets for the aortic-valve problem than for the model problem. The common subset sample sizes across both problems are $N\in\{350,100,50,10\}$, with additional subsets for the model problem. For the target tricuspid-valve application, we use all available image-derived shapes. This dataset is the most time-intensive to expand because it relies on manual segmentation of 3D TEE images.

Note that both the subset size and the dimensionality of the data space affect the dimensionality of the truncated space (Equation \ref{eq_d1_select}). For the model problem, $N > d_0 = 2$ across all subsets, whereas for the aortic- and tricuspid-valve problems, $N < d_0 = 594$. On the other hand, for the SSMs, we retain the first few modes, as discussed in Section \ref{sec_methods_ssm}. For the valves, we typically use the first three modes. The model problem is only two-dimensional, so we use the first two modes there instead.

\section{Results}\label{sec_results}

\subsection{Comparison between SSM and NUTS}\label{sec_compare_ssm_NUTS}
\begin{figure*}[htbp]  
	\centering
	\includegraphics[width=\textwidth]{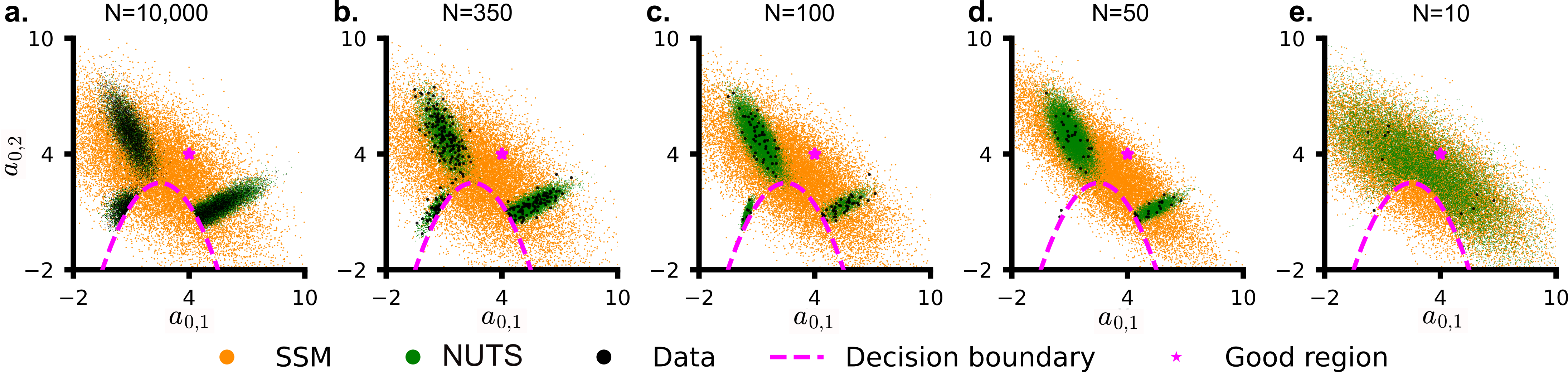}
	\caption{Synthetic data generated using SSM and NUTS given the same starting real dataset for the model problem. We varied the number of samples in the starting dataset \textbf{(a)} through \textbf{(e)}. The synthetics from NUTS capture the multimodal nature of the underlying data and a mechanism to sample inside the \textit{good} class region.}\label{fig_ssmvhmc}
\end{figure*}

We apply the SSM and NUTS methods (Section \ref{sec_threemethods}) to the model problem (Section \ref{sec_testproblems_model}). We first draw a ground-truth set of N=10,000 points and then construct nested subsets of sizes $N\in\{350,100,50,10\}$. Here, each smaller subset is drawn randomly from the next larger set. For each subset, we generate synthetic samples using both methods. Figure \ref{fig_ssmvhmc} shows the real and synthetically generated samples for each subset and method. Figure \ref{fig_ssmvhmc}(a)-(e) shows that the SSM-generated samples remain unimodal, whereas NUTS-generated samples capture the multimodal structure of the target distribution. This favorable behavior of NUTS improves when more data is included. This difference is expected: the PCA-based SSM baseline imposes a single Gaussian-like coefficient structure (Section~\ref{sec_methods_ssm}), whereas the proposed NUTS sampler uses a Gaussian mixture model prior with a data-driven number of modes (Section~\ref{sec_generatormodel}). The figure also illustrates the effect of the classifier $\mathfrak{C}$ (Section~\ref{sec:classifier}), which concentrates NUTS samples in the \textit{good} class region.  Even for the smallest subset ($N=10$, Figure~\ref{fig_ssmvhmc}(e)), where the fitted GMM effectively reduces to a single Gaussian component, the classifier term still restricts the generated samples to the good region. These results qualitatively demonstrate two advantages of NUTS over the PCA-based SSM baseline: the ability to capture multimodal shape distributions and the ability to condition generation on the good region.

\begin{figure}[htbp]  
	\centering
	\includegraphics[width=\columnwidth]{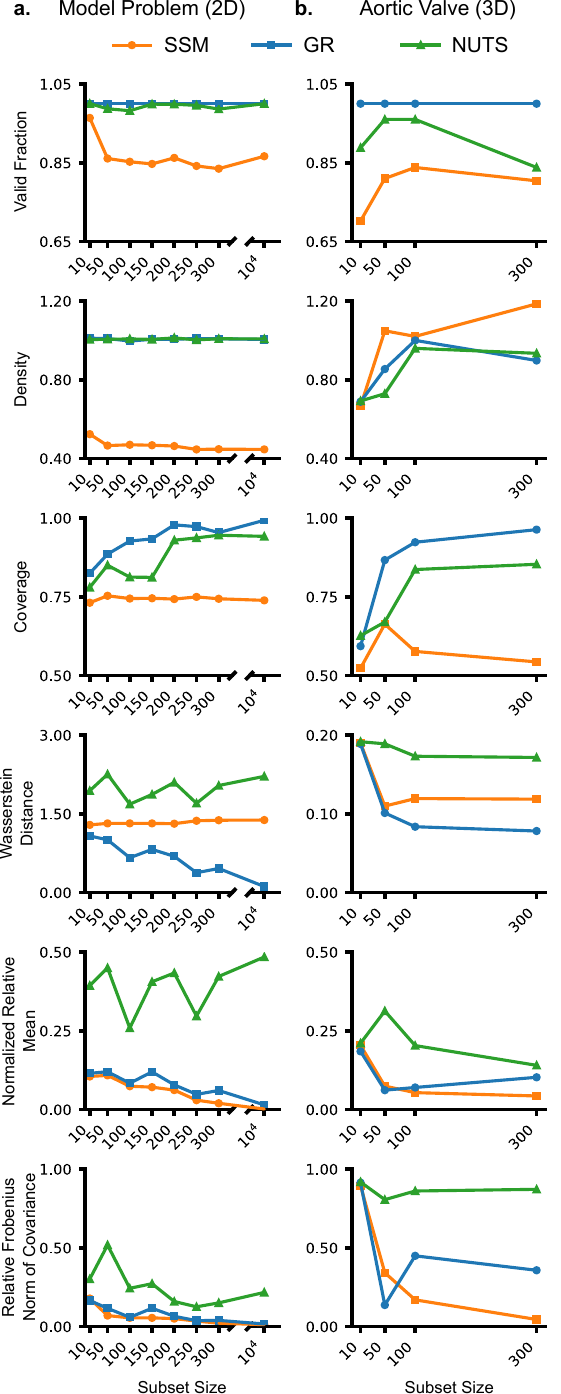}
	\caption{Various performance metrics for \textbf{(a)} the model problem (2D) and \textbf{(b)} the Aortic Valve problem (3D) as a function of the size of the starting subset drawn from the target distribution. SSM is the statistical shape method, GR is the generator-rejector method, and NUTS is the current proposed sampling strategy (Section \ref{sec_threemethods}).}\label{fig_modelvaortic}
\end{figure}
\subsection{Generated Shape Quality Metrics}\label{sec_generatedshapemetrics}

For the model and aortic-valve problems, we evaluate the SSM, GR, and NUTS methods across all available subsets. The metrics quantify both the quality of the generated samples and their relationship to the reference distribution. In statistical shape modeling, related properties are commonly evaluated through compactness, generalization, and specificity \citep{sharp_statistical_2026}. In generative modeling, similar ideas are often described in terms of fidelity and diversity. We adopt the latter terminology here and additionally define a valid-fraction metric to quantify whether generated samples lie in the \textit{good} region.

The metrics defined below capture ideas closely related to generalization and specificity. Generalization measures how well a model represents unseen geometries, while specificity measures whether generated samples remain plausible \citep{sharp_statistical_2026, munsell_evaluating_2008}. The cumulative fractional energy introduced in Eq.~\eqref{eq_fracenergy} plays a role analogous to compactness by quantifying how much variation is retained in the truncated coefficient space.

\subsubsection{Valid Fraction} \label{sec_results_validfraction}
The valid fraction $f_{\mathrm{valid}}$ is defined as the fraction of generated samples that lie in the \textit{good} region,
\begin{equation}
    f_{\mathrm{valid}} =
    \frac{1}{N_s}\sum_{i=1}^{N_s}
    \mathbb{I}\!\left[c\!\left(\mathbf{a}_1^{(s,i)}\right)=1\right],
\end{equation}
where $N_s$ is the number of generated samples and $\mathbb{I}$ is the indicator function. Computing this metric requires a mechanism for assigning class labels to generated samples. For the model problem, labels are assigned using the imposed decision boundary, so the valid fraction is computed with respect to the known ground-truth boundary. For the aortic-valve problem, no such analytical boundary is available; therefore, we use the trained classifier $\mathfrak{C}$ to estimate whether each generated shape lies in the \textit{good} region. The valid fraction for the model and aortic valve problem and its various subsets is shown in the first row of Figure \ref{fig_modelvaortic}. We see that for both problems, the valid fraction of synthetics generated by NUTS outperforms that of the SSM. We also observe that for the aortic valve, NUTS outperforms SSM significantly in the low data regimes. This improved ability to generate good shapes with less data is important for our application case of the human tricuspid valve. Because building such a dataset is costly, we want to get the most out of the data we have already collected. The GR method achieves $f_{\mathrm{valid}}=1$ by construction, since it rejects generated samples until a sample classified as \textit{good} is obtained.

\subsubsection{Fidelity and Diversity} \label{sec_results_fiddiv}
Several metric pairs have been proposed to evaluate the quality of generated samples relative to the real distribution, each with its own advantages and limitations \citep{raisa_position_2025}. Among the metrics compiled in Table 1 of \citet{raisa_position_2025}, we use density and coverage because they address known limitations of traditional precision-recall metrics while remaining widely used in generative modeling \citep{naeem_reliable_2020}. 
Density and coverage are shown in the second and third rows of Figure~\ref{fig_modelvaortic}, respectively. For the model problem, SSM underperforms NUTS and GR for both metrics. This occurs because the SSM has no mechanism to restrict samples to the \textit{good} region, whereas NUTS and GR use the classifier $\mathfrak{C}$ to condition or filter generated samples. The gap is smaller for coverage than for density because SSM samples still overlap parts of the real distribution, even though they also extend into the \textit{bad} region (Figure~\ref{fig_ssmvhmc}).

For the aortic-valve problem, NUTS and GR also achieve higher coverage than SSM. NUTS has slightly lower density than SSM, but the difference is small, particularly in the low-data subset $(N=10)$. Taken together, these results indicate that the proposed NUTS procedure generates realistic samples while covering the real distribution well.

\subsubsection{Wasserstein Distance} \label{sec_results_wasserstein}
The Wasserstein distance measures the distance between distributions in the data space, and thus requires a reference distribution for comparison. For the model problem, this reference consisted of 20,000 points drawn separately from the \textit{good} class region. For the aortic-valve problem, no analytical reference distribution is available. We therefore use the largest available subset of reference shapes $(N=350)$, which are by construction within the \textit{good} region. 
The Wasserstein distance for each method is shown in the fourth row of Figure~\ref{fig_modelvaortic}. We observe the ordering NUTS $\geq$ SSM $\geq$ GR. Interpreted alone, a larger Wasserstein distance would indicate a larger discrepancy from the reference distribution. Here, however, this trend is considered together with the higher valid fraction and coverage achieved by NUTS. Taken together, these metrics suggest that NUTS explores a broader part of the data space while still concentrating samples in the \textit{good} region, compared with SSM and GR.

\subsubsection{Mean and Covariance} \label{sec_results_meancovar}
Let $\boldsymbol{\mu}$ and $\boldsymbol{\Sigma}$ denote the mean and covariance of generated samples in the data space.  For each generated set, we compute the normalized relative mean difference $\hat{\mu}$ and the relative covariance difference $\hat{\Sigma}$ as
\begin{align}
    \hat{\mu}      &= \frac{\|\boldsymbol{\mu}_{\mathrm{syn}} - \boldsymbol{\mu}_{\mathrm{real}}\|_2}{\sqrt{\mathrm{tr}(\boldsymbol{\Sigma}_{\mathrm{real}})}}, \\
    \hat{\Sigma}   &= \frac{\|\boldsymbol{\Sigma}_{\mathrm{syn}} - \boldsymbol{\Sigma}_{\mathrm{real}}\|_F}{\|\boldsymbol{\Sigma}_{\mathrm{real}}\|_F},
\end{align}
where the subscript $\mathrm{real}$ refers to the same reference distribution used for the Wasserstein-distance calculation, and $\|\cdot\|_F$ denotes the Frobenius norm. 
The normalized mean and covariance differences are shown in the final two rows of Figure~\ref{fig_modelvaortic}. The SSM and GR methods have similar values of $\hat{\mu}$ and $\hat{\Sigma}$, whereas NUTS produces larger deviations from the reference mean and covariance. As with the Wasserstein distance, these larger deviations should not be interpreted as improved distributional agreement by themselves. Instead, when combined with the higher valid fraction and coverage reported above, they indicate that NUTS samples a broader region of the data space $\mathbf{A}_0$ while remaining within the \textit{good} region.

\subsection{Amount of Segmented Valves Required}\label{sec_amount_seg_valves}

As discussed in Sections~\ref{sec_testproblem_aortic}--\ref{sec_expt_subsets}, building valve datasets is time-intensive. This is especially true for 3D TEE datasets of adult human tricuspid valves, where each training shape requires manual segmentation. A practical question is, therefore, how many segmented shapes are needed before additional data provide diminishing returns in the quality of samples generated by the NUTS procedure. For the aortic-valve validation problem, most metrics in Figure~\ref{fig_modelvaortic} stabilize at around $N=100$ valves. We therefore use $N\approx100$ as a practical target for the number of segmented tricuspid-valve images needed to train a reliable shape generator. This estimate should not be interpreted as a strict requirement. It is based on the aortic-valve validation problem, which uses a single conditioning criterion, namely shape quality, and represents only one morphological type. Tricuspid valves, in contrast, present four distinct morphological types, or six when including subtypes \citep{hahn_proposal_2021}. Building a shape generator conditioned simultaneously on shape quality and morphological type would require sufficient examples from each anatomical type. In the current work, we therefore restrict conditional generation to shape quality alone. With larger datasets containing sufficient examples of each valve type, the same framework could also be used to sample shapes conditioned on tricuspid-valve morphology.

\subsection{Synthetically Generated Tricuspid Valves}\label{sec_syntheticgen}

\begin{figure*}[htbp]  
	\centering
	\includegraphics[width=\textwidth]{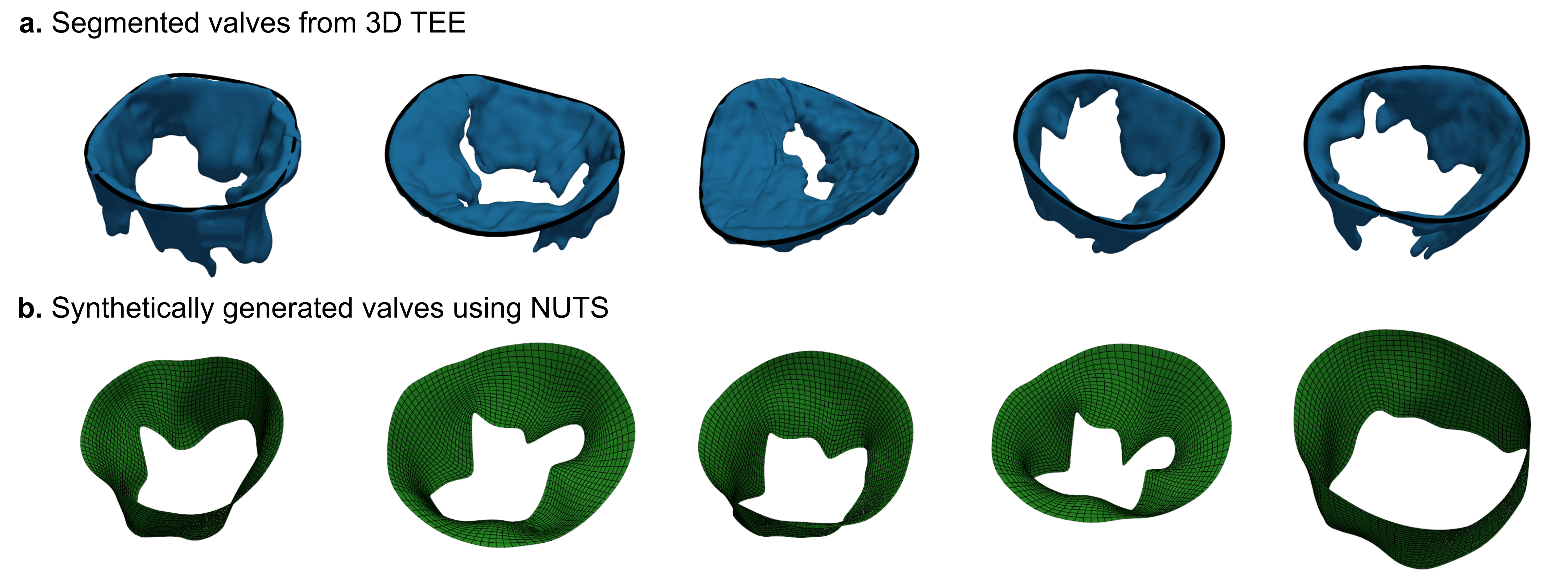}
	\caption{Real and synthetic (NUTS) tricuspid valves. \textbf{(a)} Five of the real valves and \textbf{(b)} five synthetically generated valves.}\label{fig_tv_realsynthetic}
\end{figure*}

We use the proposed NUTS method to build a synthetic shape generator for adult human tricuspid valves. The input cohort consists of ten end-diastolic tricuspid-valve shapes, whose NURBS/POD representations are obtained as described in Section~\ref{sec_testproblems_tricuspid}. We then apply the NUTS sampling procedure to generate synthetic tricuspid-valve shapes.
Figure~\ref{fig_tv_realsynthetic} shows five of the ten segmented input valves and five synthetically generated tricuspid valves. The free edges of the synthetic valves exhibit crests and troughs at similar circumferential locations. This similarity likely reflects the rotational alignment of the input segmentations along inter-leaflet boundaries (commissure). The generated synthetic valves also exhibit limited diversity, likely because the current cohort of ten valves is small relative to the practical target estimated in Section~\ref{sec_amount_seg_valves}.

\subsection{Applications of Synthetically Generated Valves}\label{sec_results_applications}

\begin{figure*}[htbp]  
	\centering
	\includegraphics[width=\textwidth]{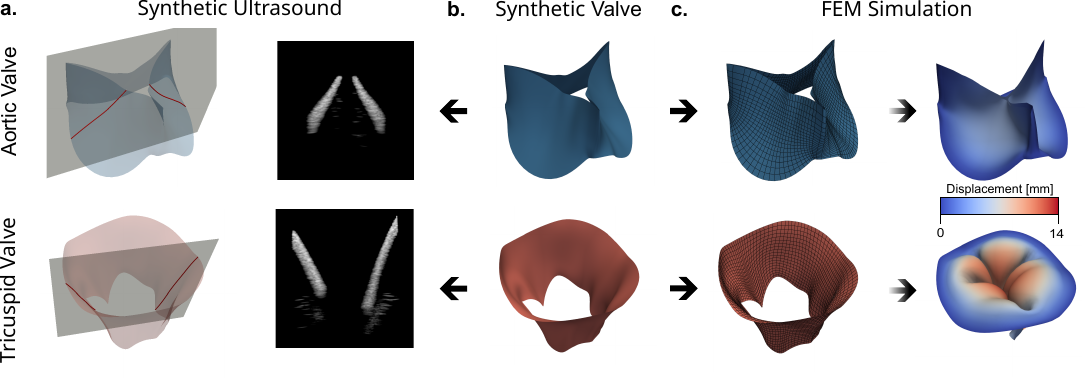}
	\caption{We generated synthetic shapes for the aortic and the tricuspid valve \textbf{(b)} using the method described in the current work. Panel \textbf{(a)} shows synthetic ultrasound images generated using these shapes and the planes for which they are generated. Panel \textbf{(c)} shows discretized reference configuration and deformed final configuration for \textit{in silico} valve mechanics simulations on these shapes. For details of synthetic image generation and \textit{in silico} model see Section \ref{sec_results_applications}.}\label{fig_tv_simsandultrasound}
\end{figure*}

The physiologically plausible synthetic valve shapes generated by the proposed framework can be used beyond shape analysis alone. We illustrate two downstream applications in Figure~\ref{fig_tv_simsandultrasound}: synthetic ultrasound image generation and finite-element valve simulation.
First, we use the generated synthetic shapes as inputs to the end-to-end differentiable ultrasound simulation framework of \citet{spencer_fully_2026}. This produces synthetic 3D ultrasound images paired with the corresponding synthetic valve geometry. Figure~\ref{fig_tv_simsandultrasound}(a) shows a two-dimensional slice through one synthetic 3D ultrasound image, together with a schematic indicating the slice location relative to the valve. 
Second, we perform finite-element simulations on the generated valve shapes shown in Figure~\ref{fig_tv_simsandultrasound}(b). The shapes are first discretized into finite-element meshes. We then apply representative physiological material properties and boundary conditions following the aortic-valve setup described in Section 3.1.2 and Figure 4(b) of \citet{dubey_graph_2026} and the tricuspid-valve setup described by \citet{mathur_texas_2022}. Several implementation details differ from these cited models: the aortic-valve mesh is coarsened near the commissures to remove near-degenerate elements, the tricuspid annulus is fixed in place, and leaflet self-contact is modeled using a penalty formulation compatible with FEBio \citep{maas_febio_2012}. These changes avoid numerical issues and allow the simulations to run in FEBio.

Figure~\ref{fig_tv_simsandultrasound}(c) shows the discretized reference configuration and the displacement field in the final pressurized configuration. Unlike the aortic valve, the tricuspid valve includes subvalvular chordae tendineae that tether the leaflets during pressurization. Because the generated synthetic shapes do not include patient-specific chordal insertion locations, we prescribe representative chordal insertions solely to demonstrate this downstream application.

\section{Discussion and Conclusion}\label{sec_discussion_conclusions}


Our study establishes a Bayesian posterior sampling method to generate synthetic shapes from a dataset of real shapes. This development was motivated by our intended application of augmenting 3D TEE datasets of adult human tricuspid valves. We verified our method using a simple model problem and then validated it on an \textit{in silico} cohort of aortic valve shapes. Finally, we applied the method for generating synthetic tricuspid valve shapes using an existing, but limited, 3D TEE dataset. 
We see qualitatively in Figure \ref{fig_ssmvhmc} and Section \ref{sec_compare_ssm_NUTS}, and quantitatively in Figure \ref{fig_modelvaortic} and Section \ref{sec_generatedshapemetrics}  that PCA-based SSM sampling is limited in two important ways. First, they cannot represent multimodal, non-Gaussian coefficient distributions when sampling from a single Gaussian-like model. Second, they do not provide a native mechanism to enforce a decision boundary or validity criterion in shape space. The proposed method for generating shapes addresses both limitations, demonstrating a crucial improvement over prior PCA-based SSMs. 
The first improvement addresses concerns about SSMs' ability to capture the underlying real distribution of the data. For example, Section 3.1 of \citep{bhalodia_deepssm_2024} highlights the issue in using a multivariate normal prior in the PCA coefficient space for shape generation. Our framework can be used there and in similar applications since the generator $\mathcal{G}$ directly addresses this issue by replacing the single Gaussian-like sampling assumption with a data-driven prior. The second improvement hinges on the use of classifier $\mathfrak{C}$. In the current work, we have used this classifier to detect geometric defects in the shape. However, this is a general mechanism that can also be used to enforce other conditions, for example, morphological type \citep{hahn_proposal_2021}. Multiple conditions could also be added and used concurrently. However, we note that additional conditions will increase the input dataset size requirement (Section \ref{sec_amount_seg_valves}). 

The proposed NUTS-based sampler consistently generates a higher proportion of \textit{good} shapes than the PCA-based SSM across all dataset sizes, even with as few as ten real shapes to start with (Figure \ref{fig_modelvaortic}). This robustness in the low-data regime is especially valuable, since annotating medical images remains expensive and time-consuming. Using the aortic valve as a case study, we find that returns from additional shapes diminish beyond roughly 100 shapes for our current model, a point well within reach even under annotation constraints. Beyond validity $(f_{valid})$, the synthetics generated are also of high quality. Relative to the PCA-based SSM, our method achieves better coverage of the shape space (Figure \ref{fig_modelvaortic}) while performing only marginally worse on density. Diversity tells a complementary story: NUTS samples sit farther from the real data's mean, covariance, and distribution (as measured by Wasserstein distance) than SSM samples, yet their valid fraction is simultaneously higher. Taken together, these results show that the proposed method explores the shape space more broadly than previous SSMs while still concentrating its samples within physiologically valid regions.

At this stage, it is worth clarifying the difference between the NUTS and GR methods, and why the former is preferred. The GR method's valid fraction, $f_{valid}=1$, is high by construction: GR simply resamples until it finds a shape that satisfies the required conditions. Coupled with density and coverage values similar to those of NUTS (Section \ref{sec_results_fiddiv}), this might make GR seem an attractive alternative. However, NUTS still outperforms GR, much as it does SSM, in its ability to sample farther into the underlying distribution while respecting these conditions, as reflected in the Wasserstein distance, mean, and covariance metrics. More importantly, NUTS explores the Bayesian posterior far more efficiently. As an adaptation of HMC (Section \ref{sec_hmc_nuts}), NUTS uses gradient information to sample more efficiently than general Markov Chain Monte Carlo (MCMC) methods \citep{betancourt_conceptual_2018}. In contrast, the efficiency of GR depends directly on the probability that an unconstrained generator sample satisfies the classifier. This acceptance probability can decrease rapidly as the valid region becomes smaller, more structured, or higher-dimensional. NUTS instead targets the posterior distribution directly and uses gradient information to explore the conditioned shape space (see Chapter 29 of \citep{mackay_information_2019}). Since the truncated POD-coefficient space in our case can be as large as 50 dimensions, this is of practical importance. Accounting for the increased sampling efficiency also makes our framework important for higher-dimensional shape representations, as any shape representation with a higher-dimensional PCA coefficient space will only widen the efficiency gap in favor of NUTS.

The synthetic shapes produced by our framework are useful beyond shape generation alone, as illustrated by the two downstream applications in Section~\ref{sec_results_applications} and Figure~\ref{fig_tv_simsandultrasound}. The first application uses synthetic shapes to generate synthetic ultrasound images, thereby producing paired synthetic images and masks. These pairs can be used to augment training datasets for any downstream task, which is especially valuable when starting from a small amount of real data. As outlined in Section \ref{sec_introduction}, a shortage of expert-annotated data is one of the challenges facing current deep-learning-based autosegmentation tools \citep{chen_recent_2022, litjens_survey_2017, tajbakhsh_embracing_2020, zhang_learning_2023}, and is especially acute for heart valves, where existing tools are trained on small datasets \citep{aly2022fully, herz_segmentation_2021}. The proposed framework provides a mechanism to efficiently bootstrap the construction of such datasets. As an example, we could expand our 3D TEE dataset of adult human tricuspid valves, for which no large-scale, openly available data currently exists, to the size recommended in Section \ref{sec_amount_seg_valves} of 100 valves. We would then use this expanded cohort to build a reliable synthetic shape generator. Finally, we would use the resulting synthetic image-mask pairs to augment the existing real image-mask pairs, thereby yielding a larger dataset for training an autosegmentation tool. Even an imperfect autosegmentation tool trained in this manner would reduce the cost of manual annotation, which would, in turn, support the creation of larger datasets and the training of more data-intensive generative models. The augmented data can be made even richer by moving from single static frames to synthetic time-resolved ultrasound spanning the cardiac cycle, from valve-open state to valve-closed state. To do so, we simulated valve closure on the synthetic shapes using the finite element method (Figure \ref{fig_tv_simsandultrasound}(c)). Passing a time-series deformation through the same ultrasound-generation framework used above would yield synthetic cine ultrasound sequences rather than single images, for a comparatively modest increase in effort.
These finite element simulations are themselves an instance of our second downstream application of \textit{in silico} modeling of valve mechanics, an area with a substantial existing literature \citep{singh2019development, laurence2021silico, mathur_texas_2022, wu_computational_2022, tondi_rt3de-based_2026} that has provided insight into valve mechanics in health, disease, and repair \citep{kunzelman1998altered, haese2024impact, zakerzadeh2017computational}. The synthetic shapes generated from our methods would be useful for expanding such investigations to span a broader set of patient geometries and conditions.

There are certain limitations to the proposed method. Because NUTS and the underlying HMC formulation it builds on (Section \ref{sec_hmc_nuts}) require derivatives of the log-probability with respect to the inputs, both the generator $\mathcal{G}$ and classifier $\mathfrak{C}$ must be differentiable, either analytically or through auto-differentiation. Therefore, the framework accommodates general generators and classifiers only insofar as they satisfy this requirement.
For the valve problems, the classifier is trained using generated shapes labeled by visual inspection of geometric defects. This labeling step scales poorly as the number of generated samples grows and is therefore a target for future improvement. At the same time, the labeling task is naturally compatible with expert input: reconstructed synthetic valves can be rendered as simple images or short view sequences and presented to clinicians or domain experts in a click-through annotation interface. This would allow experts to label geometric validity, clinical plausibility, or morphological type without interacting with the underlying coefficient space or NURBS representation.
Since these labels come from human observers, it is also worth investigating how inter-observer differences in labeling the same reconstructed shapes would affect the resulting shape labels and, in turn, the classifier itself. This is left as an open question because the current study did not examine inter-observer variability. Additionally, the energy threshold $E_T$ (Equation \ref{eq_fracenergy}) used to truncate the POD-coefficient space was set to 95\% without larger investigation. While this threshold is consistent with PCA-based SSMs, a future direction could be to make it fully data-driven \citep{mei_statistical_2008}.

Taken together, this work presents a Bayesian posterior sampling method for generating synthetic heart valve shapes. It captures the underlying distribution of real shapes more accurately than traditional PCA-based SSMs. The method also enables conditional sampling, here on geometric quality and physiological plausibility, but in principle on any criterion for which a classifier can be trained, such as valve morphological type. We further demonstrated how these synthetic shapes can bootstrap more data-hungry downstream methods, from generative image models to autosegmentation tools. This helps ease a persistent bottleneck in medical shape modeling: the scarcity of expert-annotated data. This work was motivated by, and first applied to, our growing 3D transesophageal echocardiography dataset of adult human tricuspid valves. But the underlying method is general. It readily extends to other heart valves, and to shape-modeling tasks well beyond cardiac anatomy.

\section*{CRediT authorship contribution statement}
Conceptualization: VKD, MKR, JF.
Data curation: VKD, SS, JCV, MG, FK, MKR, AP.
Formal analysis: VKD, SS, JF.
Funding acquisition: MKR, JF.
Investigation: VKD, SS, JF.
Methodology: VKD, JF.
Project administration: MKR, JF.
Resources: MKR, JF.
Software: VKD, SS, NM, CEH, LRS.
Supervision: MKR, JF, AP.
Validation: VKD, JF.
Visualization: VKD, NM, CEH, LRS.
Writing - original draft: VKD, NM, CEH.
Writing - review and editing: VKD, NM, CEH, LRS, JCV, MG, FK, IM, MKR, JF, AP.

\section*{Data and Code availability}
After the acceptance of this manuscript, all the code and data will be made available through \href{https://github.com/SoftTissueBiomechanicsLab}{this github repository} and \href{https://dataverse.tdl.org/dataverse/STBML}{this DataVerse} repository, respectively.

\section*{Declaration of competing interest}
Manuel K. Rausch reports a relationship with Edwards Lifesciences that includes: speaking and lecture fees. \

\section*{Acknowledgment}
This work was funded by NIH R01HL165251, NIH R21HL161832, U.S. NSF CMMI awards 2438943, 2235856, and 2127925 (to MKR), NIH F31HL178280 to CEH. We acknowledge additional computing support from the Texas Advanced Computing Center (Project: DMS24010). 



\appendix
\section{Details on Classifier}\label{sec_appendix_classifier}
More details for the classifier described in Section~\ref{sec:classifier} are provided here. Table~\ref{tab_classifier_arch} reports the hyperparameters for the classifier $\mathfrak{C}$ across all three test problems (Section~\ref{sec_testproblems}).

\begin{table*}[htbp]
    \centering
    \caption{Classifier architecture and training hyperparameters for the three test problems. Weight decay is implemented as $\ell_2$ regularization within the Adam optimizer.}\label{tab_classifier_arch}
    \begin{tabular}{lccc}
        \toprule
         Test Problem & Model & Aortic and Tricuspid \\
        \midrule
        Input space & $\mathbf{A}_1$ (POD coeff.) & $\mathbf{e}$ (Eq.~\ref{eq_class_ip_valve}) \\
        Input dimension $(d_1)$ & $2$ & $8$ \\
        Hidden-layer sizes & $64,\,32$ & $32,\,16$ \\
        Epochs & $3000$ & $10000$  \\
        Learning rate & $10^{-3}$ & $10^{-3}$ \\
        Weight decay & $10^{-4}$ & $10^{-4}$ \\
        Dropout probability & $0.4$ & $0.4$  \\
        Train / Val. / Test split & $0.7/0.2/0.1$ & $0.7/0.2/0.1$ \\
        \bottomrule
    \end{tabular}
\end{table*}

\section{Details on the NUTS Sampling Procedure}\label{sec_appendix_nuts}
\begin{table*}[hbtp]
    \centering
    \caption{NUTS sampling settings for the three test problems. Total samples are pooled across chains after discarding the tuning (warm-up) draws.}\label{tab_nuts_settings}
    \begin{tabular}{lccc}
        \toprule
        Test Problem & Model & Aortic and Tricuspid \\
        \midrule
        Number of chains & $10$ & $5$ \\
        Tuning (warm-up) draws per chain & $8000$ & $2000$ \\
        Retained draws per chain & $12000$ & $100$ \\
        Total pooled samples & $120{,}000$ & $500$ \\
        Chain initialization & \texttt{jitter+adapt\_diag} & SMC ($50$ particles)  \\
        \bottomrule
    \end{tabular}
\end{table*}

We provide implementation details and settings for the NUTS sampling procedure described in Section~\ref{sec_hmc_nuts}. Table~\ref{tab_nuts_settings} reports the number of chains, the number of tuning and drawing samples, and the initialization strategy for each of the three test problems (Section~\ref{sec_testproblems}). The model problem uses PyMC's default initialization strategy, \texttt{jitter+adapt\_diag}, in which the mass matrix is set to the identity and its diagonal is subsequently adapted from the variance of the tuning samples, with a uniform jitter in $\left[-1,1\right]$ added to each chain's starting point \citep{pymc2026initnuts}. The aortic- and tricuspid-valve problems instead use a sequential Monte Carlo (SMC) \citep{ching_transitional_2007} to select diverse initializations for each chain. All three problems use a target acceptance probability of $0.85$.

\bibliographystyle{elsarticle-harv} 
\bibliography{SyntheticValves.bib}




\end{document}